# Interplay Between Structural Defects and Charge Transport Dynamics in MA- and FA-Modified CsSnI₃ Thin-Film Semiconductors

Gleb V. Segal[1], Anna A. Zarudnyaya[1], Anton A. Vasilev[2], Andrey P. Morozov[1], Alexandra S. Ivanova, Lev O. Luchnikov[1], Sergey Yu. Yurchuk[2], Pavel A. Gostishchev[1*] and Danila S. Saranin[1*]

[1]LASE – Laboratory of Advanced Solar Energy, NUST MISIS, 119049 Moscow, Russia

[2]Department of semiconductor electronics and device physics, NUST MISIS, 119049 Moscow, Russia

**Corresponding authors:**

Dr. Pavel A. Gostishchev gostischev.pa@misis.ru,

Dr. Danila S. Saranin saranin.ds@misis.ru

**Abstract**

Owing high conductivity in microcrystalline thin-films, CsSnI₃ perovskite is a promising semiconductor for thermoelectrics and optoelectronics. Rapid oxidation of thin-film and intrinsic lattice strain hinders stabilization of the device performance. Cation engineering of perovskite molecule was considered as an effective strategy to tailor the structural properties and suppress the degradation processes. However, molecular engineering demands a thorough analysis of defect behavior, as it can influence ionic motion, recombination dynamics, and capacitive effects. The effective implementation of CsSnI₃ in energy conversion devices requires careful consideration of the specific properties of thin films—electrical conductivity, Seebeck coefficient, power factor, as well as electronic transients, and charge transport in the device structures. In this work, we performed a complex investigation for modified CsSnI₃ through cation substitution with methyl ammonium (MA) and formamidinium (FA). Our findings highlight a complex interplay between electrical parameters of the bare thin-films and stability of the devices (p-i-n diodes) after thermal stress. FA-CsSnI₃ showed beneficial results for stabilization under elevated temperatures with improved non-ideality factor in diode structures, enhanced shunt properties and reduced trapping. The photo-induced voltage relaxation spectroscopy performed for MA-CsSnI₃ showed relevant traps concentration of $10^{16}$ cm⁻³ with activation energy of 0.52 eV(210K) likely attributed to Sn atom defect. The obtained results are deeply analyzed and discussed.

**Introduction**

Cesium tin iodide ($CsSnI_3$) is a promising halide perovskite-based semiconductor distinguished by its high intrinsic p-type conductivity[1], in contrast to the standard lead–halide perovskites. In the black orthorhombic phase, $CsSnI_3$ exhibits a room-temperature hole concentration on the order of $10^{17}$–$10^{18}$ $cm^{-3}$ and a high hole mobility (~$10^2$ $cm^2$ $V^{-1}s^{-1}$), making it effectively a self-doped material[2]. Strong native conductivity is beneficial for energy conversion applications – it ensures efficient charge transport in photovoltaics[3] and contributes to a high-power factor in thermoelectric devices[4]. Moreover, tin-based perovskites combine decent charge transport with an ultralow lattice thermal conductivity (<1 $Wm^{-1}K^{-1}$)[5], which is a feature of the "phonon-glasses" and "electron-crystals"[6]. However, fabrication of the high-quality $CsSnI_3$ thin-films remains challenging due to lattice mismatch and internal strain resulting in affected charge transport[7–9]. To address these strain-related issues, partial substitution of the A-site cation with organic methylammonium (MA) has been explored[10,11]. Mixed-cation tin iodide compositions can subtly adjust the perovskite lattice parameters, potentially relieving lattice stress and stabilizing the high performing black phases[12]. On another hand, the defect landscape in $CsSnI_3$ is complex and markedly differs from conventional semiconductors. Intrinsic point defects typically represented with tin vacancies ($V_{Sn}$), which form under typical solution processing and acts as effective p-type dopants. As reported[13], first-principles calculations indicate that $V_{Sn}$ has a low formation energy in $CsSnI_3$ under iodine-rich, tin-poor conditions, leading to a high equilibrium concentration of acceptor defects. Self-doping in $CsSnI_3$ impacts high electrical conductivity but also correlates with structural and chemical instabilities. $CsSnI_3$ can exist in several polymorphs (black orthorhombic γ-phase at room temperature, transforming to yellow phase or amorphous phases upon degradation), and its defect states could trigger the phase transitions. In particular, the oxidation of Sn(II) to Sn(IV) is a critical degradation pathway facilitated by the defect chemistry[14,15]. In situ studies have shown that air exposure induces rapid oxidation resulting to phase transformations with vacancy-ordered double perovskite $Cs_2SnI_6$ (where all Sn is in the $4^+$ state)[16,17]. This process is essentially the filling of tin vacancies by oxygen – every $Sn^{2+}$ lost (or oxidized) creates an effective $V_{Sn}$ and two holes, which initially "self-dope" the material.

Determining quantitative defect parameters (activation energies, concentration, diffusion coefficients) is challenging for $CsSnI_3$ thin-films. Deep-level Transient Spectroscopy (DLTS) and admittance spectroscopy are widely regarded as a standard technique for the characterization of defects parameters in semiconductor-based device structures[18]. In the case of meta-stable $CsSnI_3$ specified with rapid oxidation dynamics and traps activity the analysis of numerical defect parameters is essential. To date, the DLTS and related characterization methods are more and more adapted for investigation of the halide perovskite based devices. However, the quantitative data and its interpretation of $CsSnI_3$ are lacking.

In this work, we made a complex investigation for evolution of the charge transport properties for p-i-n diodes based on modified $CsSnI_3$ films with partial substitution of A-site cations with organic methyl ammonium (MA) and formamidine (FA), which helped to stabilize the crustal lattice of solution processes samples. We analyzed the evolution of diode parameters during the dynamic transformation of $CsSnI_3$ layers and identified key differences in transient injection behavior for different cation modifications. Photoinduced Open-Circuit Voltage Spectroscopy (PIVTS) was employed, enabling the detection of defect states with an activation energy of ~0.5 eV and an estimated concentration of ~$10^{16}$ $cm^{-3}$.

## Results and discussions

In this work, two main configurations of modified $CsSnI_3$-based thin films were investigated. Briefly, A-site modified $CsSnI_3$ perovskite thin films were deposited via a solution-based spin-coating methodology. To engineer the cation composition, 20 mol% of cesium ($Cs^+$) in the perovskite structure was replaced with methylammonium ($MA^+$) or formamidinium ($FA^+$) cations to form two distinct compositions: $MA_{0.2}Cs_{0.8}SnI_3$ and $FA_{0.2}Cs_{0.8}SnI_3$. The thickness of the fabricated samples was measured using stylus profilometry, revealing a thickness of $551 \pm 15$ nm for $MA_{0.2}Cs_{0.8}SnI_3$ and $492 \pm 10$ nm for $FA_{0.2}Cs_{0.8}SnI_3$. Measurements of conductivity, Seebeck coefficient, and power factor constitute primary thermoelectric parameters, directly estimating lateral charge transport dynamics in thin-film semiconductors by quantifying carrier mobility, concentration, and energy distribution. Specifically, conductivity highlights the efficiency of in-plane charge transport and mobility, the Seebeck coefficient reveals carrier-type and energy-dependent transport mechanisms, and the derived power factor indicates the material's intrinsic thermoelectric conversion capability within the lateral transport plane. The evaluation included the simultaneous determination of both electrical conductivity σ (**fig.1(a)**) and the Seebeck coefficient α (**fig.1(b)**), employing four-probe and differential techniques, respectively. The measurements of thin films (3 × 18 mm in lateral dimensions) were performed under an ambient atmosphere in the temperature range of 300–375 K. These analyses were carried out within laboratory-made systems situated at the National University of Science and Technology MISIS (see electronic supplementary information – **ESI** for the details). The estimated errors in the Seebeck coefficient and the electrical conductivity measurements are ~8% and ~5%, respectively. Both MA– and FA– substituted $CsSnI_3$ thin films demonstrated temperature-dependent conductivity trends (300-375K) characteristic of intrinsic semiconductors, with conductivity increasing monotonically due to thermal activation. Notably, the MA–CsSnI$_3$ compound shows superior conductivity relative to the FA–modification. This can be attributed to the higher charge carrier mobility, originating from structural changes in the perovskite lattice. The smaller ionic radius of methylammonium (MA$^+$) relative to formamidinium (FA$^+$) diminishes octahedral tilting distortions in SnI$_6$ frameworks, resulting in decreased lattice strain and suppressed charge carrier localization effects. Structural analysis of CsSnI$_3$-based perovskites incorporating MA$^+$ and FA$^+$ reveals deviations of the octahedral tilt angles (α, β, γ) from the ideal 90° configuration, indicative of bond-angle distortions and lattice deformation. These angular deviations correlate with the degree of octahedral tilting, where FA$^+$-containing systems exhibit greater angular displacement compared to MA$^+$-based analogs, reflecting reduced structural symmetry and enhanced lattice distortions in FA-dominated compositions[19]. In contrast, the larger FA$^+$ cation expands lattice parameters and increases octahedral tilt angles, as confirmed by previous structural studies[20], potentially enhancing carrier scattering and diminishing overall conductivity. A positive correlation between temperature and Seebeck coefficient (Fig. 1b) was observed for both materials. MA-CsSnI$_3$ showed the gain of coefficient up to 0.78 uVcm$^{-1}$K$^{-1}$. This trend shows differences in effective mass and scattering between the two systems. Although FA-CsSnI$_3$'s larger lattice might increase effective carrier mass by reducing the band gap, lower carrier mobility reduces its Seebeck coefficient. These observations highlight the opposing effects of electronic band structure and phonon scattering on thermoelectric efficiency. The temperature dependence of the power factor (α$^2$σ), presented in (**Fig. 1c**) showed the impact of the electrical conductivity and Seebeck coefficient in the MA-based perovskite.

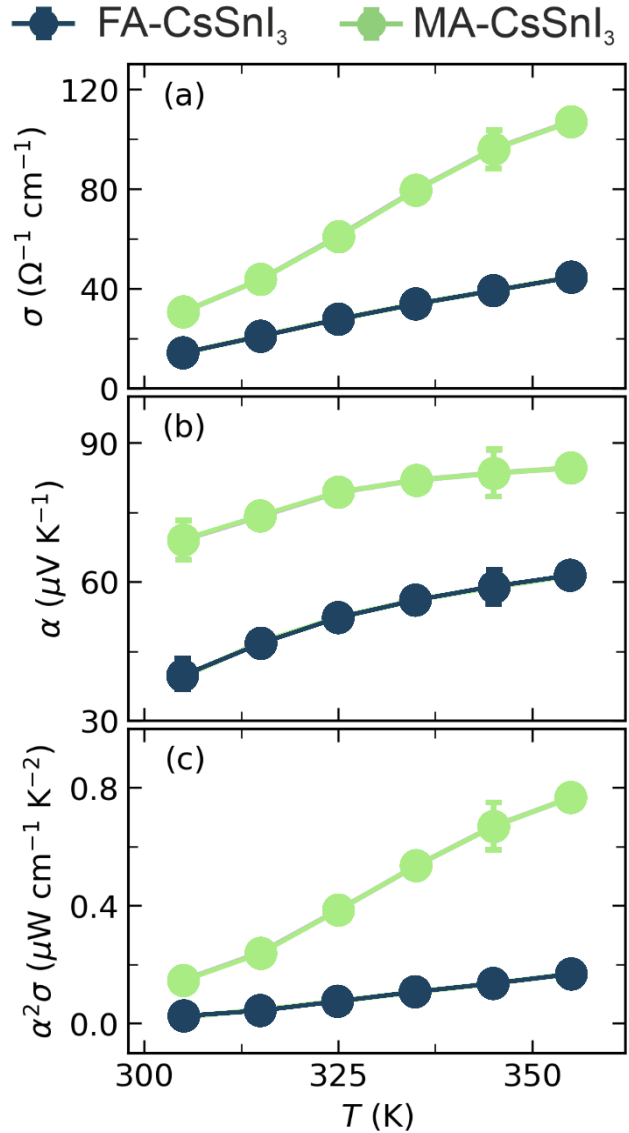

Figure 1 – Temperature dependences of (a) electrical conductivity ($\sigma$), (b) Seebeck coefficient ($\alpha$), and (c) thermoelectric power factor ($\alpha^2\sigma$) for FA-CsSnI$_3$ and MA-CsSnI$_3$ samples.

The observed data for electrical and thermic parameters was measured for lateral direction of microcrystalline thin-films with top surface positioning of the probes (see ESI for the details). The charge transport in depth of the thin-films has different specifics. Investigating electrical transport in microcrystalline thin-film CsSnI$_3$ necessitates distinguishing between lateral and vertical charge transport, as grain boundary density, defect distribution, and interface quality differ significantly across these orientations. Lateral measurements (conductivity, Seebeck coefficient, powe factor) represent in-plane carrier dynamics, while vertical configurations (for example in diode heterostructures) reveal carrier extraction efficiency and interface-dominated recombination processes. Combining lateral and vertical assessments provides comprehensive insights essential for estimation of charge transport in tailored CsSnI$_3$ thin-films.

We fabricated p-i-n structured diodes with CsSnI$_3$ modifications using the following multilayer stack (**fig.S1 in ESI**, glass/ITO (220 nm, anode)/NiO (20 nm, nanocrystalline, p-type)/CsSnI$_3$ film (MA-/FA modifications)/C60 (40 nm, amorphous, n-type)/Bathocuproine (8 nm, hole blocking interlayer)/Bi-Cu (100 nm, cathode). The detailed experimental procedure presented in ESI. To analyze the changes in the transport properties of MA- and FA- modifications of CsSnI$_3$ diodes, we measured dark volt-ampere characteristics (JV curves) (**fig.2**). The measurements were performed for the reverse and forward sweep directions in the range from 0.8 to minus 0.1 V, respectively. Both device configurations showed relevant rectification with typical diode behavior. The JV curve has 4 different regions related to the operation regimes: shunt current (I); recombination current (II); diffusion current (III) and contact resistance (IV). Fabricated p-i-n diodes exhibited minimal current (zero bias) in the range of $10^{-8}$–$10^{-9}$ A/cm$^2$. Estimation of the dark leakage current (J$_L$) at -0.1V bias revealed clear difference for the studied device configurations. For FA-CsSnI$_3$ diode, the J$_L$ was $1.1\times10^{-7}$ A/cm$^2$ (forward scan), while for reverse-mode measurements this value slightly changed to $3.2\times10^{-7}$ A/cm$^2$, displaying notable impact of hysteresis. In contrast, MA-CsSnI$_3$ samples demonstrated negligible fluctuations of JV scans, however, current leakages increased in one order up to $6.1\times10^{-6}$ A/cm$^2$. The raise of J$_L$ for MA-configuration represents the potential impact of micro pinholes or defect-rich inter-grain planes. For a pristine condition, we observed the absence of minimal current point (0 V), which is a typical feature for perovskite based diodes with NiO hole transporting layer. As reported[21,22], these effects could originate from chemical interaction between non-stochiometric nanocrystalline NiO with uncompensated Ni and organic A-site cations, such as methyl ammonium. The charge extraction properties were evaluated from the II and III regions of the dark JV. We used two-diode model[23] for fitting that allowed to extract numerical parameters non-ideality factors (*m$_1$, m$_2$*); reverse saturation currents (*J$_{01}$, J$_{02}$*); series and shunt resistances (R$_s$, R$_{sh}$), which are reported in **tab. 1**. The equivalent electric circuit (**fig.S2**) and corresponding equations (eq.S1 – S9) for calculation presented in ESI. The diode properties clearly affected both modification of CsSnI$_3$, albeit the features of rectification losses and recombination dynamics were different. First, the current leakages for MA-CsSnI$_3$ gained in almost two orders of magnitude ($2.9\times10^{-7}$ A/cm$^2$), while FA-CsSnI$_3$ showed higher resilience of J$_L$ with changes up to $1.1\times10^{-6}$ A/cm$^2$. We analyzed recombination and diffusion dark currents by estimating the non-ideality factor to determine the dominant recombination processes in the device's bulk or interfaces. In standard p-n junctions, the diffusion current is dominant at the junction when m=1, while the recombination current prevails in the space-charge region when m=2. Thus, the reduced value of m extracted for modified CsSnI$_3$ diodes could highlight the reduced impact of the interface traps and better quality of the charge transport. Typically, the halide perovskite devices don't completely follow the theory of the standard p-n junctions due to heterostructure characteristic of the diodes (p-i-n)[24–28]. Therefore, the m factor values could exceed the range between 1 and 2. Reports on various halide perovskite-based devices show m-values exceeding 2, which require modifications to the equivalent electrical circuits for better description of critical features such as multiple heterojunctions, trapping, and ionic defects. For the pristine conditions, MA-CsSnI$_3$ diodes had m=2.055 which slightly extended conventional range of the values. The significant changes were observed for FA-CsSnI$_3$, which showed m=1.403 with clear reduction of the recombination impact. Dark saturation current (J$_0$) in diode is determined by recombination current. The comparison of the extracted J$_0$ values for MA- and FA- devices revealed the showed the decrease of both J$_{01}$, J$_{02}$ for more than order of magnitude (up to 10$^{-5}$ A/cm$^2$). Analysis of the shunt resistance correlated to the results of J$_{min}$, J$_0$. For R$_{sh}$ was 7.4 kOhm*cm$^2$, while for FA-CsSnI$_3$ enhanced to 12.4 kOhm*cm$^2$. Despite improved conductivity of FA-CsSnI$_3$, the series resistance slightly decreased from 5.2 to 3.7 Ohm*cm$^2$ compared to MA-CsSnI$_3$.

Migration of the ionic defects is a critical feature of the halide-perovskite based devices[22,29,30]. Accumulation of the charged defect at the interfaces induces the electrochemical corrosion at the interfaces[31,32], possible phase transformation[33,34] and decomposition[35,36]. We measured forward and reverse JV scans (**fig.S3**, **ESI**) to estimate the impact of hysteresis. Both device configuration demonstrated negligible fluctuations with the sweeps and highlighted the reduced contribution of the ionic motion for pristine conditions of the samples.

To estimate the stability of the rectification properties for the modification of CsSnI$_3$, we used a rapid aging treatment – the dark storage of the devices in the oven at 80 °C for 24 hours. The application of thermal stress to the diode samples induced the gain of hysteresis effect in JV curves (**fig.S4**, **ESI**). MA-based devices highlight a critical loss in shunt properties with reduction of R$_{sh}$ to 200 Ohm*cm$^2$. FA-CsSnI$_3$ had stronger rectification resilience with negligible changes in R$_{sh}$. On the other hand, the series resistance for FA-CsSnI$_3$ was affected with an increase to 46.5 Ohm*cm$^2$. Interestingly, that thermal stress induced the reduction of the m-values for both device configurations. After thermal stress, the m for MA-CsSnI$_3$ was 0.933 and for FA-CsSnI3 was 1.211. An m-value below 1 indicates a fundamental change in device physics. The methylammonium cation is known to be thermally unstable, and at 80°C, significant degradation likely occurred at the perovskite/transport layer interfaces. The extremely low m-value (<1) represents a case where diode models become inadequate due to the introduction of parallel conduction paths. This masked recombination effects due to overwhelming non-diode currents. In contrast, FA-CsSnI$_3$ showed a more moderate reduction in m-value (from 1.403 to 1.211) with preserved shunt resistance properties.

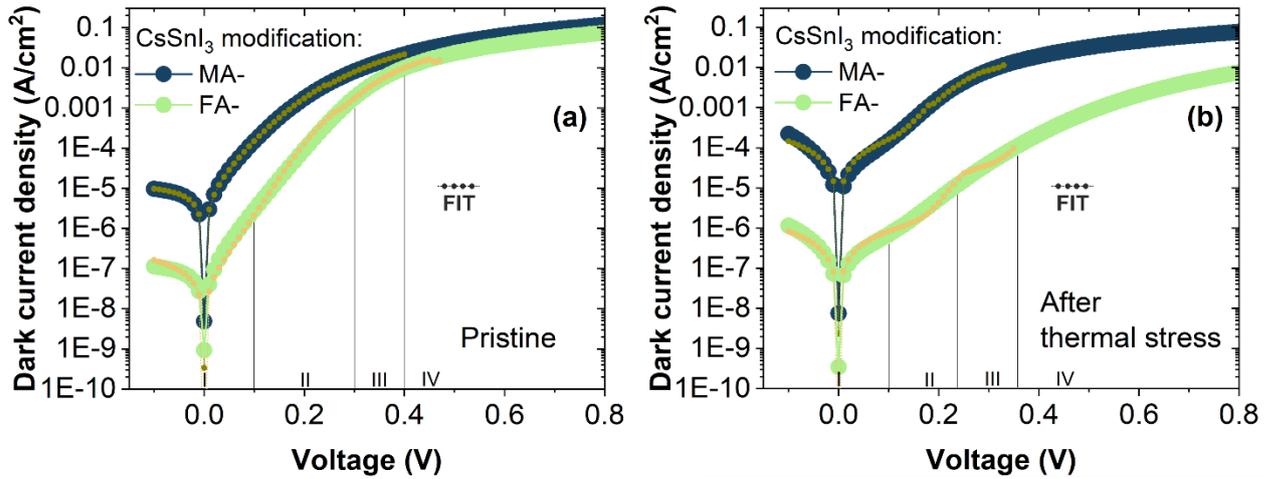

Figure 2 – The dark JV curves for p-i-n diodes with MA and FA modifications of CsSnI$_3$ measured for pristine conditions (a), and after thermal stress (b)

Table 1. The extracted parameters of 2-diode model for the dark JV curves of CsSnI$_3$ diodes

| Device configuration | $m_1$ | $m_2$ | m ($m_1+m_2$) | $J_{01}$, A*cm$^{-2}$ | $J_{02}$, A*cm$^{-2}$ | $R_{sh}$, ohm*cm$^2$ | $R_{s}$, ohm*cm$^2$ |
|---|---|---|---|---|---|---|---|
| **Pristine conditions** | | | | | | | |
| MA-CsSnI$_3$ | 1.249 | 0.806 | 2.055 | 5.6E-6 | 9.2E-4 | 7420 | 3.94 |
| FA-CsSnI$_3$ | 0.794 | 0.609 | 1.403 | 7.1E-9 | 4.3E-5 | 12450 | 5.25 |
| **After thermal stress** | | | | | | | |
| MA-CsSnI$_3$ | 0.604 | 0.329 | 0.933 | 1.9E-8 | 1.85E-4 | 200 | 5.77 |
| FA-CsSnI$_3$ | 0.691 | 0.520 | 1.211 | 4.3E-11 | 4.32E-5 | 12200 | 46.5 |

In parallel, we did a shelf-life test in dark storage condition for 250 hours. The changes in the rectification properties differed from the thermal stress conditions (**fig.S5, ESI**). FA-CsSnI$_3$ diode showed negligible fluctuations in the dark current and represented relevant stability and reproducibility of the JV curve. For MA-CsSnI$_3$ devices we observed the slight increase of J$_L$ up to $1 \times 10^{-5}$ A/cm$^2$. The comparison of shelf-life and thermal stress results revealed that distinct changes in CsSnI$_3$ device were triggered at elevated temperatures. We performed additional analysis of the transient process, which revealed the difference of transport properties for MA- and FA- based devices in the range from 0 to 80°C.

To evaluate the recombination dynamics of the diode strictures we usen transient current measurements (**TCM**) with high resolution in time. Halide perovskites have a vast response (micro and nanosecond range), particularly, thin-film CsSnI$_3$ owns short lifetimes of charge carries in the range of 10 ns. The intrinsic defect in halide perovskite could be presented with ionic species

(iodine vacancies, uncompensated A—site cations, iodine interstitial, etc.) and introduce the slow component to the current transients[37–39]. We performed TCM at 100 kHz frequency with a forward applied bias of 1V using precise oscilloscope system (see ESI for the details). Transient processes in $CsSn_3$-based diodes vary with temperature due to the thermal activation of defect states, changes in carrier mobility, and alterations in recombination rates. At lower temperatures, traps with deeper energy levels dominate, causing slower transients due to reduced thermal energy and prolonged carrier capture/release; conversely, at higher temperatures, increased thermal excitation accelerates carrier dynamics, resulting in faster transient responses and changes in recombination pathways. In our work we used the temperature range between 0 to 80°C for the extraction of corresponding rise and fall times ($\tau_R$, $\tau_F$), corresponding to the 10 and 90% of the saturation signal. **Fig. 3** illustrates the temperature dependence of these parameters; measurements were conducted at constant temperatures to ensure accuracy (see **ESI** for the details). For the rise regime, we observed the sub-microsecond response. In both regimes of TCM, $MA-CsSnI_3$ modification showed an increased transient time, compared to $FA-CsSnI_3$. MA- and FA-modification showed a difference of approximately 60 ns at 0°C, which gradually reduced to ~10 ns at maximum heating conditions. As expected, the maximum $\tau_R$ was measured at 0°C with values 0.2-0.3 us for the studied diode samples and decreased in two times (0.12-0.15 us) at 80°C. For the fall regime, the response times ($\tau_F$) increased to the range of 0.6-0.7 μs, but the average difference in the average values reduced to 20 ns at low temperatures (0 to 20°C). Notably, at high temperature of TCMs (80°C), the $\tau_F$ values were very close for both $CsSnI_3$ modifications (0.43-0.44 μs). The form of the response signal had rapid growth and slow component prior saturation for both regimes of the operation, indicating the difference in recombination dynamics of the samples with variation of the organic cations[39,40].

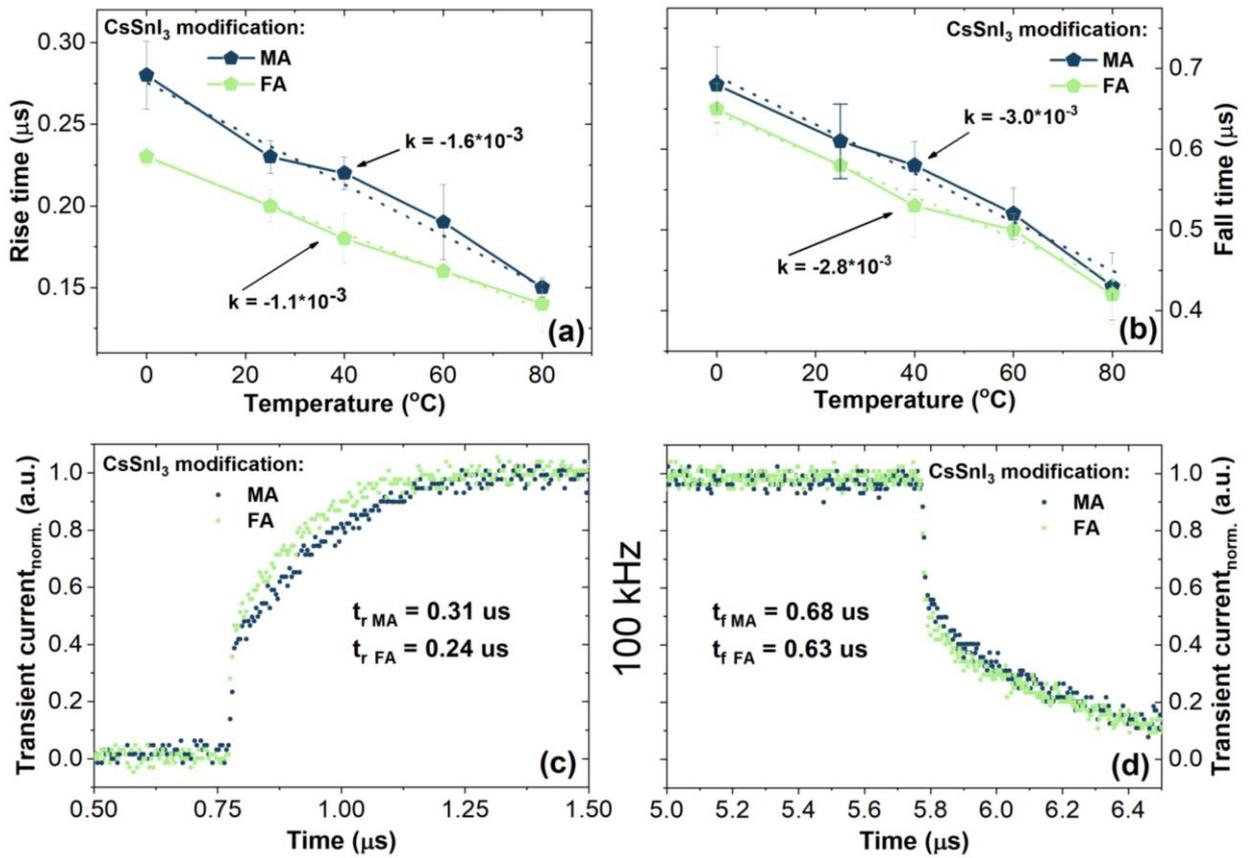

Figure 3 – The extracted $\tau_R$ (a) and $\tau_F$ (b) values for the TCMs in the temperature range of 0-80°C, representative transient plots for rise (c) and fall (d) regimes at 0°C

The observed data represents a complex interplay of conductivity and charge carrier processes for the studied thin-film configurations of $CsSnI_3$. Fast rise times imply that capacitive components (decreased depletion width) respond immediately; traps either filled rapidly or uninvolved in the injection[41,42]. Despite higher conductivity of $MA-CsSnI_3$, we observed a more pronounced slow component in the transient response, attributable to defect-mediated ionic motion and trap-assisted processes. In contrast, FA substitution exhibited suppressed defect-driven slow processes, resulting in faster transients. At elevated temperatures, ionic and trap kinetics were accelerated in both materials. This behavior could result from the capacitance values of the studied films, as t=RC. The load resistance and the total capacitance (including the device's junction capacitance) determine the RC time constant.

The frequency-dependent capacitance of $MA-CsSnI_3$ and $FA-CsSnI_3$ diode structures was measured at 0 V bias and 298.6 K (**fig.4**). At low frequency (0.1 kHz), the capacitance values ($C_0$) were 14.9 nF for $MA-CsSnI_3$ and 12.1 nF for $FA-CsSnI_3$. Both devices exhibited a capacitance plateau in the 10–100 kHz range, with values ranging from 13.5 to 12.6 nF for $MA-CsSnI_3$ and from 10.5 to 9.8 nF for $FA-CsSnI_3$. At higher frequencies, a sharp drop in capacitance was observed, attributed to the inability of the interface charge to respond to rapid voltage oscillations. The cutoff frequencies were 340 kHz for MA-CsSnI$_3$ and 220 kHz for FA-CsSnI$_3$, indicating superior high-frequency response in the MA-based device. The reduced capacitance in FA-based devices can be attributed to structural factors, such as larger lattice parameters induced by the FA$^+$ cation, which influence charge carrier dynamics. However, as the temperature increased, the MA-based diodes exhibited a steeper reduction in both $\tau_R$ and $\tau_F$. This improvement is consistent with the conductivity trends shown in Fig.1(a), where the MA- configuration displayed a more pronounced increase in electrical conductivity with temperature.

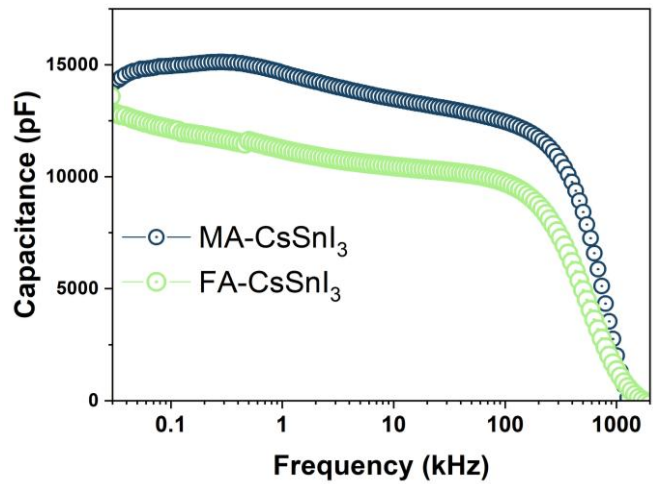

Figure 4 – Capacitance – frequency plot for the p-i-n devices with MA- and FA- modifications of CsSnI$_3$ thin-films

At low forward bias, the traps concentration was estimated to be approximately $5 \times 10^{16}$ cm$^{-3}$. The observed high-frequency cutoff (**fig.S6**, ESI) is attributed to the voltage drop across the series resistance (R$_s$).

The photo-induced voltage spectroscopy (PIVTS) method[43] enables the investigation of mobile ion kinetics by analyzing open-circuit voltage (V$_{oc}$) relaxation following illumination. In this experiment, a 470 nm LED with an optical power density of approximately 250 mW/cm² was used. The V$_{oc}$ relaxation was recorded after switching off the light source (**fig.5(a)**). By conducting these measurements at varying temperatures, the influence of mobile ions can be revealed, allowing for the estimation of their diffusion coefficient (**fig.5(b)**). In systems where the built-in electric field of the p-i-n structure is modulated by ionic charge distribution[43], the characteristic time τ required for an ion to travel a distance equivalent to the Debye length (L$_D$) is given by время $L_D^2 / D = \tau$, where D is the diffusion coefficient (see **eq. (1)**). By correlating the rate of V$_{oc}$ relaxation with this time constant, it becomes possible to extract the diffusion parameters of mobile ions.

$$\tau = L_D^2/D = \frac{\epsilon\epsilon_0 k_B T}{q^2 N_i D_0} \exp\left(\frac{E_A}{k_B T}\right) \longrightarrow D(T) = \frac{\epsilon\epsilon_0 k_B T}{q^2 N_i \tau} \qquad (1)$$

Where $\epsilon_0$ – permittivity of free space, $\epsilon$ – permittivity of perovskite, $k_B$ – Boltzmann constant, $T$ – temperature, $q$ – electric charge of an electron, $N_i$ – concentration of charges, $D_0$ – diffusion coefficient, $E_A$ – activation energy of diffusion

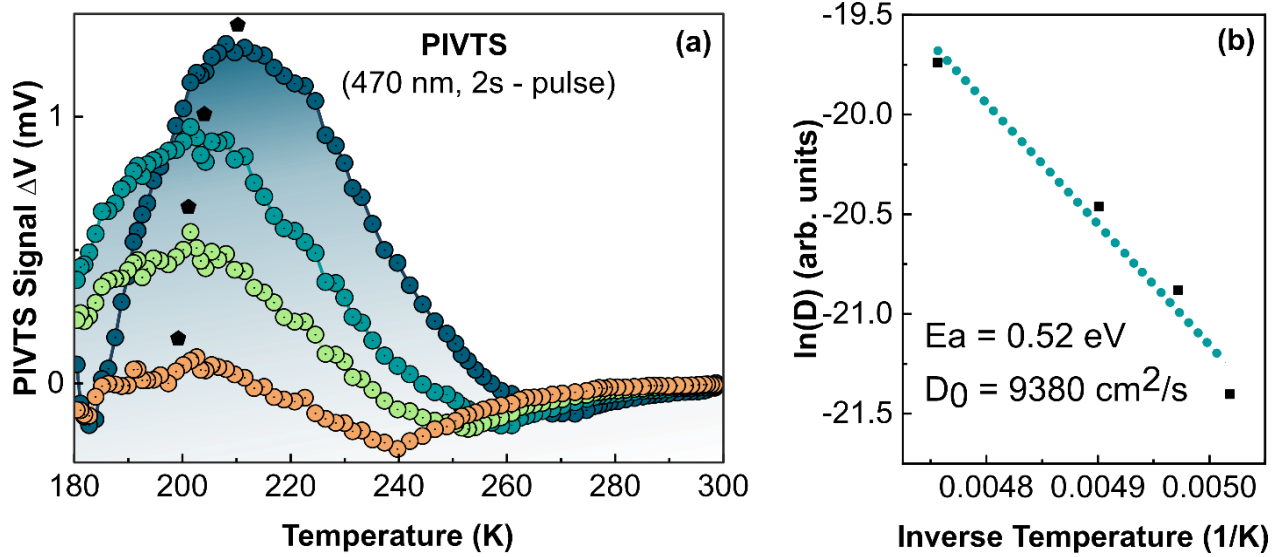

Figure 5 – PIVTS spectra of the MA-CsSnI$_3$ diodes under optical pumping (a), Arrhenius plot (b)

The calculated trap density ($\sim 5 \times 10^{16}$ cm$^{-3}$) in MA-CsSnI$_3$ film is on the order of magnitude of typical solution-processed lead perovskites ($10^{15}$–$10^{16}$ cm$^{-3}$) and aligns with other investigations on CsSnI$_3$[44,45]. Notably, it is comparable to trap densities in amorphous silicon ($\sim 10^{16}$ cm$^{-3}$ dangling-bond defects)[46], underscoring that further passivation is needed to reach the defect levels of more developed inorganic thin-film technologies, like CdTe, CIGS (N$_{traps}$ $\sim 10^{13}$–$10^{14}$ cm$^{-3}$)[47,48]. A deep trap at 0.52 eV in CsSnI$_3$ could be associated with donor-type defects like tin interstitials (Sn$_i$) or halide-related defects can create deep levels ($\sim 0.5$ eV from a band edge) that trap electrons[49]. Recently, Yu et al. reported high temperature [50] activated defects with Ea$\sim 0.6$ eV (410K) and 0.4 EV (350K), representing an ionic species of undercoordinated Sn$^{2+}$. So, we assume that in or case, the deep centers are related to the defective position of metal cation. D$^0$ $\sim 10^4$ cm$^2$/s suggests fast ionic transport; such a high value implies very mobile defects/ions (perhaps due to a high concentration of migration-friendly vacancies) and is above diffusion coefficients in typical nano/microcrystalline thin-film systems and could be potentially affected via shunting pathways. This highlights the need for combined strategies of A-site modification with inter-grain passivation and antioxidants incorporation for compensation of the ionic species (SnF$_2$, thiosemicarbazide, etc.)[51–53].

## Conclusions

In this work, complex transport properties of modified $CsSnI_3$ thin-film semiconductors ($MA_{0.2}Cs_{0.8}SnI_3$, $FA_{0.2}Cs_{0.8}SnI_3$) were investigated under thermal stress (80 °C, 24h) and shelf-life conditions (250h). Temperature-dependent electrical conductivity ($\sigma$) analyses in the range of 300–375 K showed a superior conductivity of $MA$-$CsSnI_3$, attributed to decreased lattice distortions and enhanced carrier mobility compared to $FA$-based samples. Comprehensive measurements yielded $\sigma$ values increasing monotonically with temperature (up to ~0.78 $\mu V \cdot cm^{-1} \cdot K^{-1}$ Seebeck coefficient $\alpha$ for $MA$-$CsSnI_3$), emphasizing thermoelectric potential. Detailed diode characterization via JV measurements revealed minimal leakage current ($J_L$) for $FA$-$CsSnI_3$ (~$10^{-7}$ $A/cm^2$) compared to higher values (~$10^{-6}$ $A/cm^2$) for $MA$-based films, highlighting pinhole or inter-grain defects as critical conduction pathways. Analysis of diode parameters indicated significant recombination reduction in $FA$-$CsSnI_3$ (ideality factor m=1.403) relative to $MA$-$CsSnI3$ (m=2.055). Rise times in transient current measurements (0–80 °C) fell from ~0.3 $\mu s$ to ~0.15 $\mu s$, highlighting faster ionic kinetics and defect processes at higher temperatures. Additionally, PIVTS analysis allowed the extraction of defect parameters, revealing a deep trap state at ~0.52 eV, potentially associated with tin interstitials, with estimated trap density around $5 \times 10^{16}$ $cm^{-3}$ and a high ionic diffusion coefficient ($D_0$ ~$10^4$ $cm^2/s$). Comprehensive transport analysis and recombination dynamics demonstrated crucial interplay between conductivity, ionic mobility, defect states, and diode properties, providing significant insights for practical $CsSnI_3$ applications in thermoelectric converters and stable thin-film devices.

The interplay between structural defects and charge transport dynamics in $MA$- and $FA$-modified $CsSnI_3$ thin-films originates from the contrasting roles of lattice distortions and defect-induced recombination pathways: while $MA$ substitution enhances electrical conductivity due to reduced lattice distortions and improved carrier mobility, it simultaneously introduces inter-grain defects increasing leakage currents and recombination, whereas $FA$ substitution mitigates recombination losses by reducing defect density, despite exhibiting comparatively lower conductivity. This balance highlights the critical trade-off between carrier transport efficiency and defect-driven recombination in tailoring $CsSnI_3$ semiconductors for optimized device performance.

**Declaration of competing interest**

The authors declare that they have no known competing financial interests or personal relationships that could have appeared to influence the work reported in this paper.

**Data availability**

Data will be available on request.

**Acknowledgments**

The authors gratefully acknowledge the financial support from the Russian Science Foundation with project № 22-79-10326.

**The Electronic Supplementary Information (ESI) for the paper:**

**Interplay Between Structural Defects and Charge Transport Dynamics in MA- and FA-Modified CsSnI₃ Thin-Film Semiconductors**

**Gleb V. Segal[1], Anna A. Zarudnyaya[1], Anton A. Vasilev[2], Andrey P. Morozov[1], Alexandra S. Ivanova, Lev O. Luchnikov[1], Sergey Yu. Yurchuk[2], Pavel A. Gostishchev[1*] and Danila S. Saranin[1*]**

[1]LASE – Laboratory of Advanced Solar Energy, NUST MISIS, 119049 Moscow, Russia

[2]Department of semiconductor electronics and device physics, NUST MISIS, 119049 Moscow, Russia

**Corresponding authors:**

Dr. Pavel A. Gostishchev gostischev.pa@misis.ru,

Dr. Danila S. Saranin saranin.ds@misis.ru

**Experimental section:**

*Materials*

All organic solvents – dimethylformamide (DMF), dimethyl sulfoxide (DMSO), isopropyl alcohol (IPA), chlorobenzene (CB) were purchased in anhydrous, ultra-pure grade from Sigma Aldrich, and used as received. 2-Methoxyethanol was purchased from Acros Organics (99.5+%, for analysis), $HNO_3$ (70%). Devices were fabricated on $In_2O_3$: $SnO_2$ (ITO) coated glass ($R_{sheet}$<7 Ohm/sq) from Zhuhai Kaivo company (China). $NiCl_2 \cdot 6H_2O$ (from ReaktivTorg 99+% purity) used for HTM fabrication. Tin iodide ($SnI_2$, >99.9%), Cesium iodide (99.99%), trace metals basis from LLC Lanhit, Russia and Methylammonium iodide (MAI, >99.99% purity from GreatcellSolar), Formamidinium iodide (FAI, >99.99% purity from GreatcellSolar), were used for perovskite ink. Fullerene-C60 (C60, >99.5%+) was purchased from MST NANO (Russia). Bathocuproine (BCP, >99.8% sublimed grade) was purchased from Osilla Inc. (UK).

*Inks preparation*

 A nickel oxide precursor ink was prepared by dissolving $NiCl_2 \cdot 6H_2O$ powder in 2-methoxyethanol (2-ME) at a concentration of 35 mg/mL. To 1 mL of this solution, 20 μL of 70% nitric acid ($HNO_3$) was added. A 1.3 M perovskite precursor solution, comprising organic cations (OC) such as methylammonium (MA) and formamidinium (FA), was prepared by mixing OCl:CsI:SnI₂ in a 0.2:0.8:1 ratio. All perovskite precursor solutions were dissolved in a solvent mixture of dimethylformamide (DMF) and dimethyl sulfoxide (DMSO) in a 4:1 ratio. The solutions were stirred for 30 minutes and subsequently filtered before deposition. For the hole-blocking layer, a BCP solution was prepared by dissolving BCP in isopropyl alcohol (IPA) at a concentration of 0.5 mg/mL and stirring overnight at 50°C.

### Sample preparation

*Thin-film samples*

Thin-film samples for electrical conductivity, Seebeck coefficient, and thermoelectric power factor were fabricated on a glass (Soda Lime 1.1 mm) with a structure glass/perovskite.

*Devices*

**p-i-n** diodes for dark JV, transient current measurements, frequency-dependent capacitance and PIVTS measurements were fabricated with inverted planar architecture ITO coated glass/c-NiO /perovskite/C60/BCP/Bi-Cu.

*Layers deposition routes*

Firstly, the ITO substrates were cleaned with acetone and IPA in an ultrasonic bath and activated under UV-ozone irradiation for 30 min. The NiO HTL was spin-coated in an air atmosphere with a relative humidity not exceeding 40% with the following ramp: (1 s – 500 rpm, 10 s – 4000 rpm). The deposited NiO layer was annealed at 125 °C for 15 minutes and at 310 °C for 1 hour. Perovskite films were crystallized on top of HTL with a solvent-engineering method. The deposition and crystallization processes of perovskite layers were conducted inside glove box with an inert nitrogen atmosphere. Perovskite precursors were spin-coated with the following ramp: step 1: 10 sec, 1000 rpm. Step 2: 30 sec, 4000 rpm. Antisolvent CB with volume of 200 µl was dropped on substrate at 25 sec of the second step. Then, substrates were annealed at 50 °C (20 min) and 100 °C (20 min) for conversation into the black perovskite phase. As an ETL 40 nm of C60 was deposited with the thermal evaporation method at $10^{-6}$ Torr vacuum level. The BCP layer was spin-coated at 4000 rpm (30 s) and annealed at 50 °C (1 min) inside a glove box. Metal cathode Bi (15 nm) and Cu (85 nm) thick was also deposited with thermal evaporation through a shadow mask to form a 0.15 cm$^2$ active area for each pixel.

### Characterization

Dark JV curves Keithley 2400 SMU in 4-wire mode and a settling time of $10^{-2}$ s. The time-resolved transient current measurements was measured with DIGILENT Analog Discovery Pro 3450 (2 units), which were used as oscilloscope and pulse generator.

Electrical conductivity σ and the Seebeck coefficient α, were measured employing four-probe and differential techniques, respectively. The measurements of the thin films in geometry (3 × 18 × 1 mm$^3$) were performed under an ambient atmosphere in the temperature range of 300–375 K. Analyses were carried out within laboratory-made systems situated at the National University of Science and Technology MISIS. The estimated errors in the Seebeck coefficient and the electrical conductivity measurements are ~8% and ~5%, respectively.

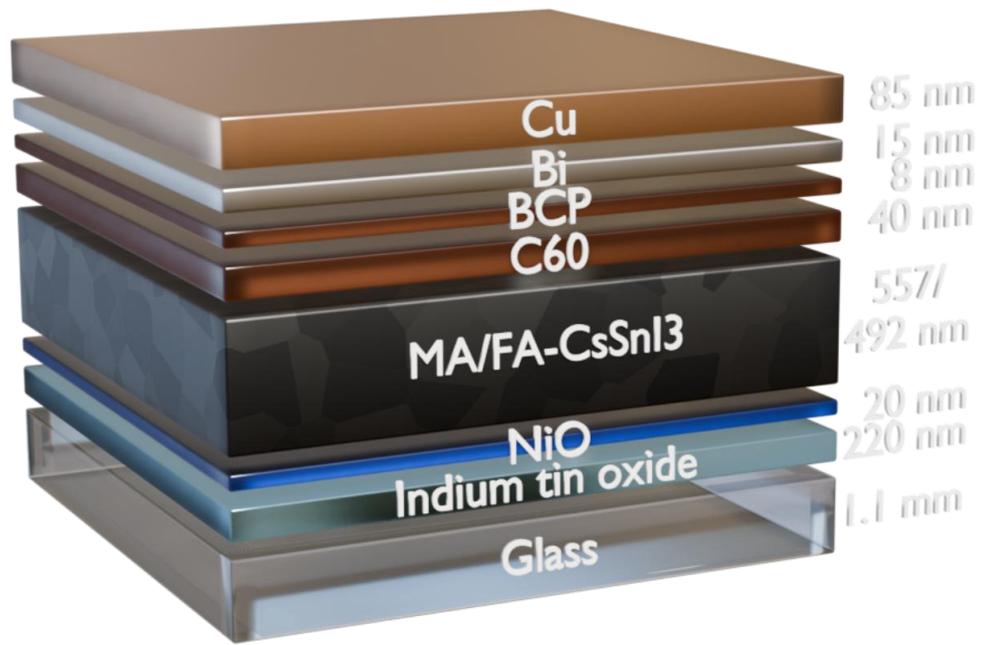

Figure S1 – The schematics of p-i-n diodes fabricated with CsSnI$_3$ modifications

## 2- diode model fitting

The equation for the diode current-voltage curve, including series (Rs) and shunt (Rsh) resistance, has the following expression (S1):

$$J = J_0 \cdot \left( \exp\left( \frac{q \left( V - J \cdot R_s \right)}{m \cdot k \cdot T} \right) - 1 \right) + \frac{V - J \cdot R_s}{R_{sh}} .$$

(S1)

For common pn junction, such a model describes the characteristics of real structures quite accurately. However, for a p-i-n structure, such a model does not always allow us to express the characteristic corresponding to experimental results. The reason for this is the presence of two barriers from the p- and n-regions, therefore, to calculate the current-voltage characteristics of the p-i-n structure, we used a two-diode model, represented by the equivalent circuit in Fig. S2. [2]

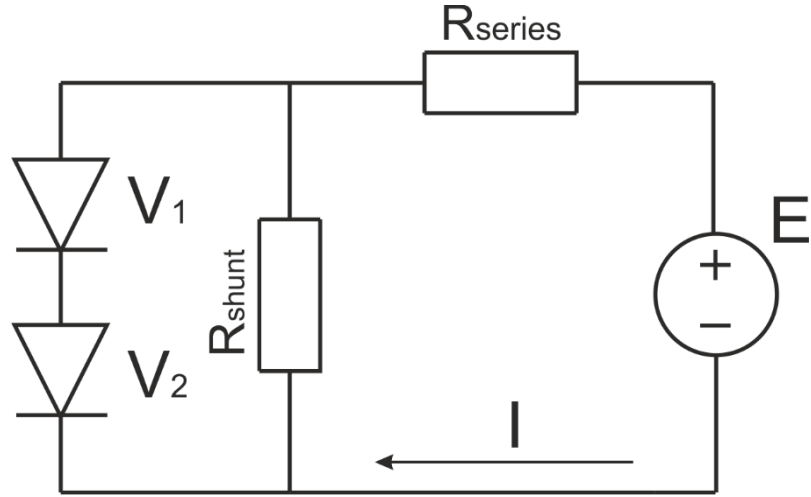

Figure S2 – Double diode circuit for modeling of pin PSC

Diodes 1 and 2 in this circuit are ideal diodes, the current-voltage characteristics of which are described by the expressions S2-S3:

$$J_1 = J_{01} \cdot \left( \exp\left( \frac{qV_1}{m_1 \cdot k \cdot T} \right) - 1 \right)$$

(S3)

J2=J02·(exp(qV2m2·k·T)−1)J2=J02·exp⁡𝑗₀qV2m2·k·T−1 (S4)

Two diode structures are connected in series, so the currents are equal to each other.

$$J_d = J_{01} \cdot \left( \exp\left( \frac{qV_1}{m_1 \cdot k \cdot T} \right) - 1 \right) = J_{02} \cdot \left( \exp\left( \frac{qV_2}{m_2 \cdot k \cdot T} \right) - 1 \right)$$

(S5)

Since the equivalent circuit is branched, it is necessary to solve a system of equations obtained from Kirchhoff's laws to calculate the current-voltage characteristic, For a given voltage V, unknown values are the currents in the diode and shunt resistance circuits $J_d$ and Rsh. The voltages applied to each diode $V_1$ and V2, and the total current J, which we must find for each given voltage. The system of equations (S4) - (S7) is sufficient for the numerical calculation of the current – voltage characteristics according to the two-diode model, but it is necessary to calculate the model parameters: leakage currents of each individual diode $J_{01}$ and $J_{02}$, non-ideality coefficients $m_1$ and $m_2$ of each diode, series resistance $Rs$ and shunt resistance $Rsh$.

$$V_1 + V_2 = V - J \cdot R_s$$

(S6)

$$J_{R_{sh}} = \frac{V - J \cdot R_s}{R_{sh}}$$

(S7)

$$J = J_d + J_{R_{sh}}.$$

(S8)

The calculation was done with one of the methods for multi-parameter optimization, in which the objective function requiring minimization of the sum for the differences between the experimental and theoretically calculated currents (S8).

$$\sum_{i=1}^{m} (J_{exi} - J_{teori})^2$$

(S9)

where m is the number of experimental points.

Calculation program was developed in Borland Delphi 7, which allows determining the parameters of a two-diode structure using multi-parameter optimization by the coordinate descent method.

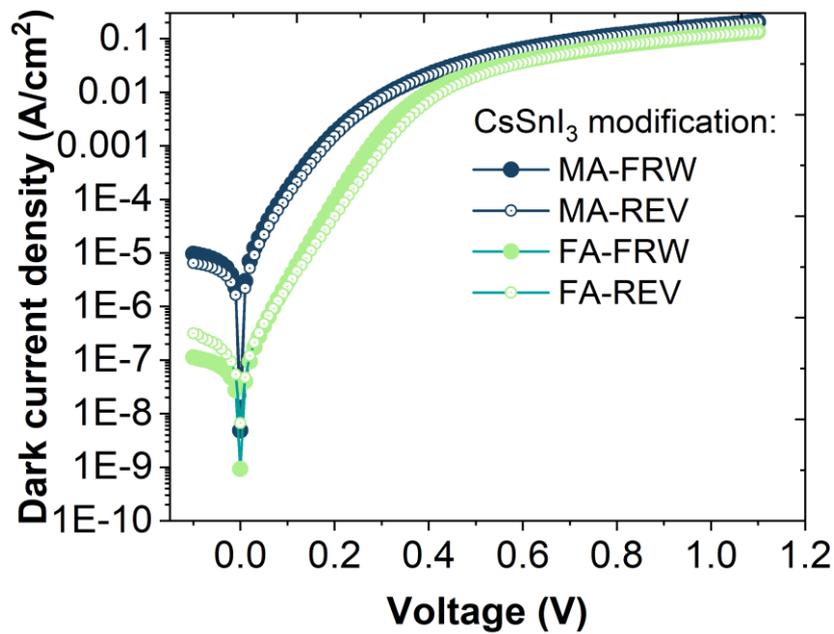

Figure S3 – The reverse and forward scans of dark JV curves for MA- and FA- CsSnI$_3$ modifications for pristine conditions

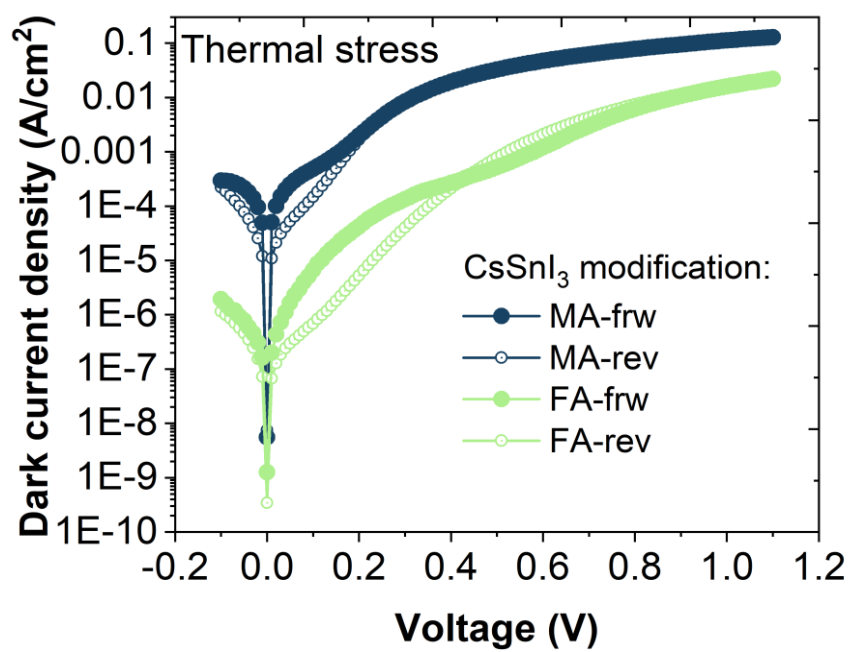

Figure S4 – The reverse and forward scans of dark JV curves for MA- and FA- CsSnI$_3$ modifications after thermal stress

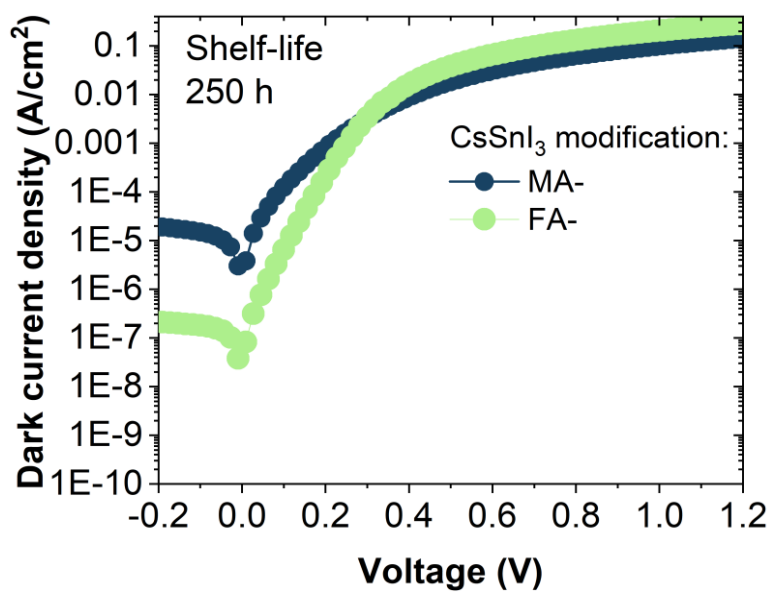

Figure S5 - The dark JV curves for MA- and FA- CsSnI$_3$ modifications shelf life test

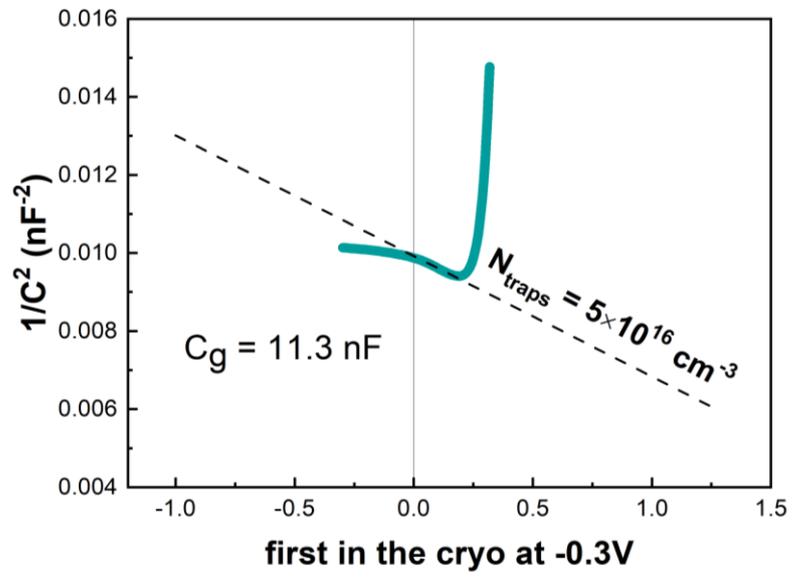

Figure S6 – The Extraction of Ntrap via capacitance measurements